\documentclass[reprint, floatfix, superscriptaddress, aps, pre]{revtex4-1}
\usepackage{amsmath,amssymb}
\usepackage{graphicx,hyperref,color,bm}
\graphicspath{{./}{figures/}}

\begin{document}

\title{Run-and-Tumble Escape in Pursuit-Evasion Dynamics of Intelligent Active Particles}
\author{Segun Goh}
\affiliation{Theoretical Physics of Living Matter, Institute for Advanced Simulation, Forschungszentrum J{\"u}lich, 52425 J{\"u}lich, Germany}
\affiliation{School of Liberal Studies, Sejong University, 05006 Seoul, Korea}
\author{Dennis Haustein}
\affiliation{Theoretical Physics of Living Matter, Institute for Advanced Simulation, Forschungszentrum J{\"u}lich, 52425 J{\"u}lich, Germany}
\affiliation{Institute of Biological Physics, Universit\"at zu K\"oln, 50937 K\"oln, Germany}
\author{Gerhard Gompper}
\email{g.gompper@fz-juelich.de}
\affiliation{Theoretical Physics of Living Matter, Institute for Advanced Simulation, Forschungszentrum J{\"u}lich, 52425 J{\"u}lich, Germany}
\affiliation{Institute of Biological Physics, Universit\"at zu K\"oln, 50937 K\"oln, Germany}
\date{\today}
\begin{abstract}
The pursuit-evasion game is studied for two adversarial active agents, modelled as a deterministic self-steering pursuer and a stochastic, cognitive evader. The pursuer chases the evader by reorienting its propulsion direction with limited maneuverability, while the evader escapes by executing sharp, unpredictable turns, whose timing and direction the pursuer cannot anticipate. To make the target responsive and agile when the threat level is high, the tumbling frequency is set to increase with decreasing distance from the pursuer; furthermore, the range of preferred tumbling directions is varied. Numerical simulations of such a pursuit-target pair in two spatial dimensions reveal two important scenarios. For dominant pursuers, the evader is compelled to adopt a high-risk strategy that allows the pursuer to approach closely before the evader executes a potentially game-changing backward maneuver to pull away from the pursuer. Otherwise, a strategy where the evader tumbles forward with continuous slight adjustments of the propulsion direction can significantly increase the capture time by preventing the pursuer from aligning with the target propulsion direction, while maintaining the persistence of the target motion. Our results can guide the design of bioinspired robotic systems with efficient evasion capabilities.
\end{abstract}

\maketitle

\section{Introduction}

Since the Cambrian explosion~\cite{mars:06}, animals have developed diverse morphologies 
and a large variety of motility modes. In the course of animal evolution, the development of 
sensory organs, e.g., the acquisition of vision, has enhanced the precision of sensing, 
thereby facilitating sophisticated predation schemes~\cite{ferr:10,weis:19}. 
Consequent competition among species implies high evolutionary selection 
pressure~\cite{park:98,aber:12}, the so-called arms race hypothesis~\cite{verm:94}, which 
might in turn have contributed to the robust motility of animals by enabling them either to 
chase prey or to escape from predators. 
This suggests that the capability to sense the position of other animals, 
to process this information, and to act accordingly, is critical not only at 
the individual level but also for coordinating competitive interactions between predator and prey. 
Accordingly, several pursuit and evasion strategies have been developed, which can 
ultimately affect population dynamics under predation pressure~\cite{sinc:03}. 
Here, a good strategy needs to take into account several aspects of pursuit dynamics, 
e.g., energy management~\cite{zhan:19}, group chasing \cite{ange:12,jano:17,meye:23}, vision 
ranges~\cite{osha:09} as well as noise~\cite{furu:02,bern:22,su:23}.

The pursuit-evasion game has also been a central theme in robotics~\cite{chun:11}. 
Despite the seemingly straightforward analogy with animal behavior, implementing effective 
pursuit-evasion strategies for robots remains challenging. In contrast to animals equipped 
with neural intelligence, a robot can only adapt its actions through a programmed algorithm 
that makes decisions based on sensory inputs. Although  machine learning and artificial 
intelligence facilitate the optimization of pursuit strategies~\cite{gonu:24,bajc:24,tang:25,xiao:25}, 
model-based, control-theoretic algorithms~\cite{mitc:05,zhou:24m,zhou:24p,rao:25} continue to 
provide valuable insights. From this perspective, efficient and adaptable strategies for 
autonomous robots can benefit from nature-inspired predation scenarios.

From a physical point of view, the disparity in maximum speed, maneuverability, and endurance of 
pursuer and target plays an important role in the outcome of a pursuit~\cite{wils:13,wils:18}. 
One of the widely investigated 
strategies in this direction is ``hopping'', or ``turning gambit''. In many animal-predation scenarios, 
predators are more athletic, for instance faster and heavier, than prey~\cite{wils:18}, while prey is smaller, 
usually more agile, capable of performing rapid 
maneuvers, and can therefore make sharp, unpredictable turns~\cite{hump:70,wils:15,li:17}. 
Here, randomness of motion is an important aspect of escape from a more athlectic pursuer, as
it hampers the predictability of future target locations, and thus the possible anticipation 
of target trajectories and early adaptation of pursuer motion. However, simple random-walk-like
motion of the target is also inefficient, as it only leads to a highly localized diffusive
behavior.
In the case of rodents, it has been suggested that bipedal hopping may indeed be associated 
with predator avoidance~\cite{moor:17, dome:11}. Examples of such maneuvers further include 
ricochetal movement in primates~\cite{crom:07}, jumping in hares~\cite{kuzn:17}, C-start 
escape in fish~\cite{gazz:12}, turning gambit in birds~\cite{hede:01}, as well as directed 
turning in invertebrates~\cite{muij:14}. 
Remarkably, also in robotics, adversarial learning may lead to an agile escape strategy~\cite{pint:17,tang:20}.

Accordingly, we explore a turning-based maneuver in pursuit-evasion games as an interesting escape strategy.
Instead of considering the mechanics and dynamics of individual turning events, we focus in our
study on the stochastic decision making, global dynamics and statistical properties of 
run-and-tumble escape. To do so, we employ models of active matter physics, where both pursuers 
and evaders in pursuit dynamics are described as self-propelled active agents, which are best 
characterized by their speed and persistence of motion. For the recognition of other agents, 
these models have to be augmented by sensing, information processing, and adaptation of motion capabilities. 
For the pursuer, we employ a minimal cognitive model with sensing of the instantaneous position of the 
target~\cite{barb:16,goh:22,negi:22}. For the evader, the irregular zigzagging motion is conveniently 
modeled by a run-and-tumble dynamics, with straight 
run episodes interrupted by random tumbling events. Thus, we employ a run-and-tumble particle (RTP) -- 
originally proposed as a model for the stochastic motion of bacteria, such as \emph{E. coli} 
\cite{schn:92,berg:04} in search for food -- augmented by state-dependent tumbling rates 
controlled by sensing and information processing of the instantaneous pursuer location.

Employing concepts from statistical physics, we aim to elucidate the interplay between 
long-term stochastic dynamics and evasion outcomes, instead of focusing on the detailed 
trajectory of a single turning epoch, as already done, e.g., in the classical study of 
Ref.~\cite{howl:74}, where optimal predator-avoidance strategies for prey are discussed. 
We mainly address two questions. First, when should the evader respond to the  
pursuit by tumbling? In particular, we revisit the faster-start hypothesis~\cite{walk:05}, 
which states that prey should start their escape maneuver as quickly as possible. 
Second, what is a promising direction for an evader to select for successful escape? An 
obvious guess is that an evader should move in the direction pointing away from the pursuer. 
Remarkably, however,  counter-intuitive escape responses toward the threat are observed in a 
significant number of cases, up to 10-50\% of time~\cite{dome:11}, in animal predation. 
Our analysis predicts that the timing and the direction of turning are strongly interdependent 
and, therefore, need to be regulated simultaneously for an optimal evasion outcome.

\section{Model}
\label{sec:model}

We consider a pursuit scenario, where a pursuer at position ${\bm r}_p$ chases a target 
at position ${\bm r}_t$. The pursuer is modeled as a self-steering active particle, a deterministic 
version of the intelligent active Brownian particles~\cite{goh:22, negi:22}.
The particle is assumed to move with constant speed $v_p$ and aims to redirect its motion
toward the target, so that the dynamics is governed by
\begin{align}
\dot{\bm r}_p &= v_p {\bm e}_p, \label{eq:iABP_r}\\
\dot{\bm e}_p &= C_p {\bm e}_p\times ({\bm e}_p \times {\bm r}/r). \label{eq:iABP_e}
\end{align}
Here ${\bm e}_p$ and  ${\bm r} = {\bm r}_t - {\bm r}_p$ denote the propulsion direction of the 
pursuer and the vector connecting the target and pursuer positions, respectively, and 
$r \equiv |{\bm r}|$ the distance. The essential feature of the self-steering of the self-propulsion 
direction ${\bm e}_p$ is that the pursuer reorients toward the target direction ${\bm r}$ with 
a limited steering torque determined by the parameter $C_p$. 

As the model for the evader, we propose an intelligent run-and-tumble particle (iRTP), which 
is a conventional run-and-tumble particle (RTP)~\cite{tail:08,solo:15.2}, augmented by 
sensing-controlled tumbling rates and tumbling angles -- depending on the information about 
instantaneous position of the pursuer. An iRTP, as a standard RTP, moves in 
a consecutive sequence of ``runs'' at constant speed $v_t$ governed by
\begin{align}
\dot{\bm r}_t = v_t {\bm e}_t, \label{eq:iRTP_r}
\end{align}
interrupted by random ``tumble'' events with a maximum rate $C_t$. For the iRTP, however, the 
tumbling rate is assumed to depend on the distance to the pursuer, in order to reduce the risk of capture. 
Also the tumbling direction can be anisotropic to enhance the chance of evasion. Thus, the dynamics 
of the propulsion direction of an iRTP is generally governed by the master equation
\begin{align} \label{eq:target_RTP}
&\frac{dP({\bm e}_t|{\bm r};t)}{dt} = 
  C_t\, \omega (r) 
 \left\{ \mathcal{W}[P({\bm e}_t|{\bm r};t)]- P({\bm e}_t|{\bm r};t)\right\},
\end{align}
where the function $\omega(r)$ and the functional $\mathcal{W}$ model the state-dependent 
tumbling rate and tumbling direction, respectively.

\begin{figure}
    \centering
    \includegraphics[width= 0.9\columnwidth]{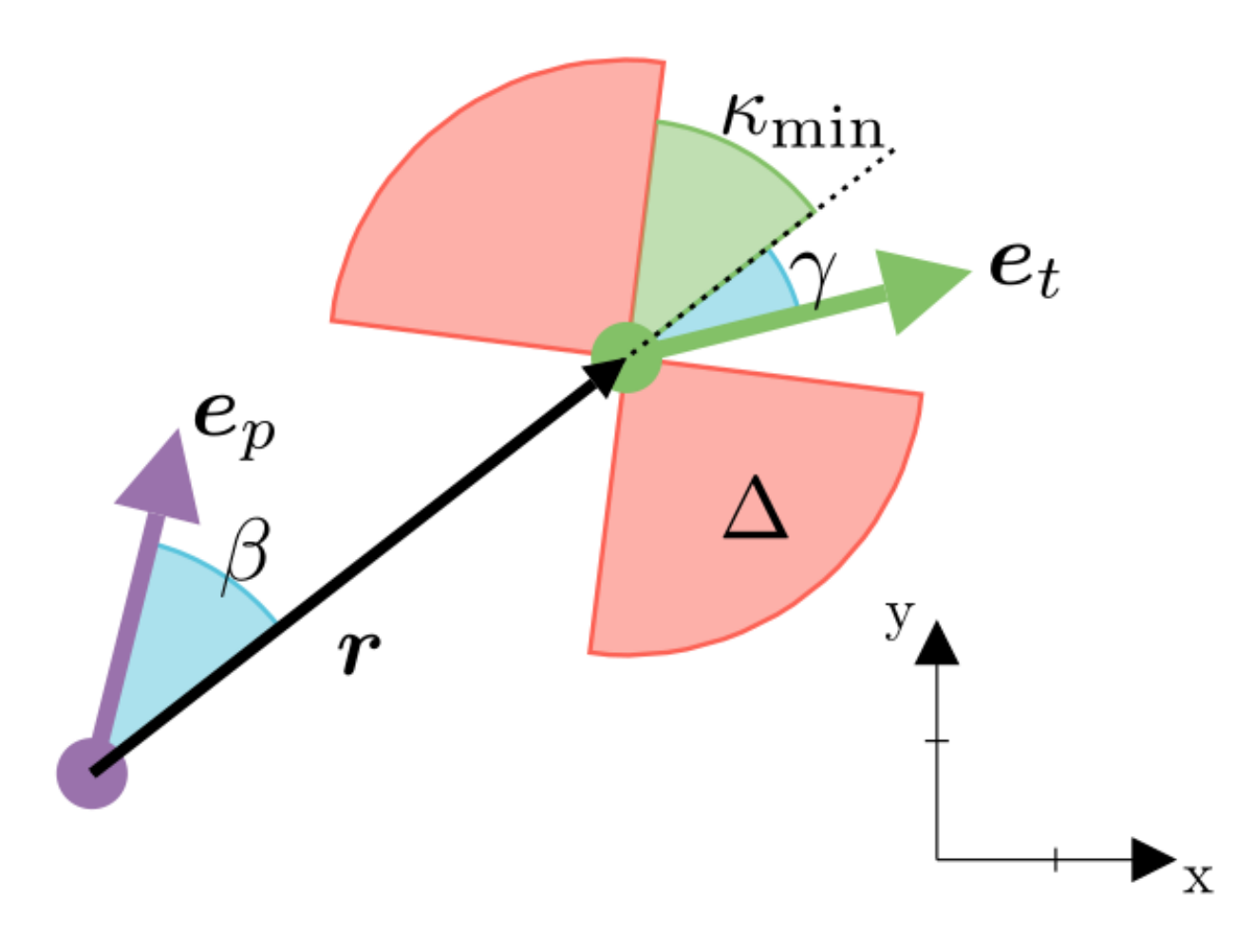}
    \caption{\label{fig:model}
    Illustration of a pursuer-target pair in two spatial dimensions, with the bearing angle $\beta$ 
    and the escape angle $\gamma$. The pursuer (purple) with propulsion direction 
    ${\bm e}_p = (\cos{\phi_p}, \sin{\phi_p})^T$ and the target (green) with 
    ${\bm e}_t = (\cos{\phi_t}, \sin{\phi_t})^T$ are located at the origin and at ${\bm r}$, respectively. 
    The two symmetric ranges of random tumbling angles between $\kappa_{\rm min}$ and $\kappa_{\rm min}+\Delta$  
    on the left- and right-hand sides of the target direction ${\bm r}$ are represented by the shaded (red) circular sectors.}
\end{figure}

We then focus on surface-bound pursuit, i.e., motion in a two-dimensional planar system,
compare \textbf{Figure~\ref{fig:model}}. In polar coordinates, the distance vector is 
${\bm r} \equiv r(\cos{\theta}, \sin{\theta})^T$, and the self-propulsion directions of pursuer and 
target are written as ${\bm e}_p=(\cos{\phi_p}, \sin{\phi_p})^T$, and ${\bm e}_t = (\cos{\phi_t},\sin{\phi_t})^T$, 
respectively. As an additional (external) parameter, we introduce the contact distance $\sigma$, 
which implies that the target and pursuer are in contact when $r \leq \sigma$.
In the following, length and time are measured in units of $\sigma$ and $\sigma/v_t$, respectively. 
Accordingly, we introduce two dimensionless parameters, which 
are the target agility and the pursuer maneuverability,
\begin{align}
\Omega_t \equiv \frac{C_t \sigma}{v_t}, \quad \Omega_p = \frac{C_p \sigma}{v_p},
\end{align}
respectively. The maneuverability $\Omega_p$ also corresponds to the inverse of the 
minimal turning radius of the pursuer measured in units of $\sigma$~\cite{goh:22,gass:23}. Such a concept 
of maneuverability has also been employed to 
model drifts due to inertia in the context of differential games~\cite{isaa:65}.
Now self-steering is given as a function of the bearing angle $\beta \equiv \theta -\phi_p$ 
with $-\pi \leq \beta < \pi$, together with the escape angle 
$\gamma \equiv \phi_t - \theta$ with $-\pi \leq \gamma < \pi$, where the latter
indicates the target escape direction with respect to the pursuer direction, in the form
\begin{equation}
\label{eq:beta}
\dot{\beta} = \frac{\sigma}{r}(\alpha \sin{\beta} + \sin{\gamma}) - \alpha \Omega_p \sin{\beta}, \end{equation}
and $\alpha \equiv v_p/v_t$ is the speed ratio between the pursuer and target.
The self-steering may not reduce the bearing angle for distances smaller than the turning radius $r/\sigma < 1/\Omega_p$~\cite{goh:22}, as the target cannot be approached within the small distance $r$ due to a limited maneuverability $\Omega_p$. From Equation~\eqref{eq:iABP_r} and~\eqref{eq:iRTP_r}, the time evolution of $r$ in polar coordinates 
is obtained as 
\begin{align}
\label{eq:r} 
\frac{\dot{r}}{v_t} = - \alpha \cos{\beta} + \cos{\gamma}, 
\end{align}
which indicates that the distance dynamics is governed by 
the bearing angle $\beta$ and the escape angle $\gamma$, capturing the self-steering of the pursuer 
and of the target, respectively.

\begin{figure*}
    \centering
    \includegraphics[width= 0.9\textwidth]{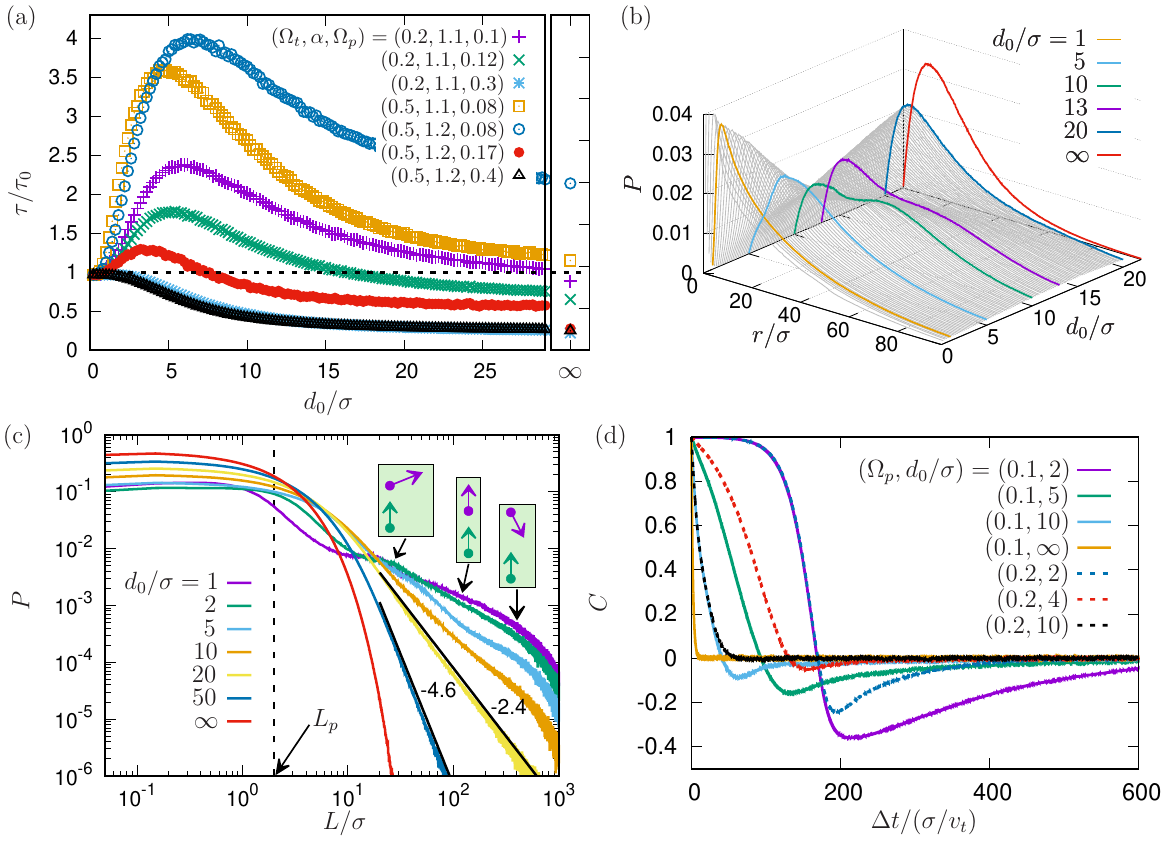}
    \caption{\label{fig:isotropic}
    Isotropic tumbling. 
    (a) Capture time as a function of the alert distance $d_0$ for various combinations of 
    $\Omega_t$, $\alpha$, and $\Omega_p$, as indicated. 
    (b) Distance and (c) run-length distributions for $\Omega_t=0.5$, $\Omega_p=0.091$, $\alpha=1.1$. 
    Grey lines in (b) represent distributions at values indicated on the $d_0$ axis with an increment of $0.2$. 
    A few curves exhibiting typical behaviors are highlighted in color. Notice the log-log scale employed 
    for the run-lengths distributions. Black solid lines in (c) show power-law distributions, with indicated 
    exponents. Typical configurations depicted in insets are extracted from the $d_0=2$ data. The persistence 
    length for the distance-independent case ($d_0 \to \infty$) given as $L_p/\sigma = 1/\Omega_t = 2$ is 
    also indicated.
    (d) Orientational-autocorrelation-function decays for various combinations of $\Omega_p$ and $d_0$, as 
    indicated. Results shown are for $\Omega_t = 0.5$, $\alpha = 1.1$, and $r_{\rm init}/\sigma =20$.
    }
\end{figure*}

To specify the escape strategy, $\omega(r)$ and $\mathcal{W}$ in Equation~\eqref{eq:target_RTP} have to be defined. 
When a predator is very far away, there is no urgent need for prey to tumble -- in fact, a change of the 
direction of motion may even be detrimental for survival, as it may imply a reduction of the distance to 
the predator. As a predator comes closer, the risk of capture by the predator increases and the prey has 
to initiate a evasive maneuver, i.e., has to enhance the tumbling rate. Accordingly, we consider an exponential 
dependence of the tumbling rate on the predator distance $r$,
\begin{align}
\omega (r) = e^{-r / d_0},
\end{align} 
where $d_0$ is a characteristic alert distance. We note that $\omega(r) \leq 1$ for all $r\geq 0$, so 
that tumbling is largest for small distances $r$, and decreases with increasing $r$. In the limit 
$d_0\to \infty$, the distance to the pursuer becomes negligible and $\omega = 1$ is recovered. As will 
be demonstrated below, adjustable tumbling rates can enhance the escape performance of the prey 
significantly -- an expression of the obvious fact that the information about the presence and 
proximity of a predator is crucial for escape. The adaptation of the tumbling angle is captured 
by the functional $\mathcal{W}$. 
We assume a parity symmetry, such that the probability to tumble left and right are equal, 
while the escape angle is limited to the range from $\kappa_{\rm min}$ to 
$\kappa_{\rm max} \equiv \kappa_{\rm min} + \Delta$, see Figure~\ref{fig:model} for illustration. 
Thus, the functional $\mathcal{W}$ is
\begin{align} \label{eq:tumbling_angle_W}
\mathcal{W}[P(\gamma,\phi_t;t)] =
\frac{1}{2\Delta}& \left[ \int_{\theta+ \kappa_{\rm min}}^{\theta + \kappa_{\rm max}} {\rm d}\phi_t'\, P(\phi_t',\gamma;t) \right. \nonumber \\
&\left. + \int_{\theta -\kappa_{\rm max}}^{\theta-\kappa_{\rm min}} {\rm d}\phi_t'\, P(\phi_t',\gamma;t) \right].
\end{align}
Since both $\gamma$ and $\phi_t$ are involved via $\theta$, 
it is not straightforward to analyze the coupled dynamics of the pursuer and target. Only in the special case of preferred forward tumbling, an approximate analytical solution can be obtained, see Section~\ref{sec:forward}.
For a ``dumb'' target, with constant tumbling rate ($\omega=1$) and isotropic 
tumbling direction, the target dynamics is decoupled from that of the pursuer. Then, the run time -- the time 
of persistent forward motion -- is $\tau_p = 1/C_t$, and the corresponding run-length $L_p = v_t/C_t =\sigma/\Omega_t$.  

We first discuss a few limiting cases. If $\alpha < 1$, i.e. a larger target-than-pursuer 
speed, the target can always escape from the pursuer, as discussed in the pursuit of mammalian 
predators~\cite{hirt:20}. On the other hand, if $\alpha \gg 1$, the target self-propulsion 
speed becomes irrelevant. Accordingly, we consider only $1< \alpha \leq 1.3$ in this study. 
Our previous study~\cite{goh:22} showed that $\alpha \approx 1.6$ is the limit below which 
the target self-propulsion becomes relevant for pursuit on straight target trajectories.

For the target agility $\Omega_t$, two trivial limits exist. For 
$\Omega_t \ll 1$, the tumbling of the target is negligible, and therefore, the 
pursuit dynamics is ballistic. In this case, $\alpha$ is the dominant factor determining 
the outcome of pursuit. For $\Omega_t \gg 1$, escape is 
again inefficient as $\tau_p \to 0$, and the target motion is essentially diffusive around its initial 
position, which provides no advantage compared to a stationary target. In between these two limits, 
we expect nontrivial behavior, with no obvious advantage for either the pursuer or the target.

\section{Isotropic tumbling}
\label{sec:iso_tumbling}

We first consider the case where a target tumbles isotropically in all directions, i.e., 
$\kappa_{\rm min}=0$ and $\kappa_{\rm max} = \pi$, with $\Delta = \pi$.

\subsection{Capture time}

The capture time $\tau$ is defined as the first-passage time between an 
initial configuration with ${\bm e}_t = {\bm e}_p$ (aligned propulsion directions) and 
${\bm r} = - r_{\rm init} {\bm e}_p$, where  $r_{\rm init}$ denotes the initial separation, and a 
final configuration, where the distance between the 
target and pursuer become smaller than the contact distance $\sigma$, i.e.,
the pursuer is assumed to capture the target as soon as they come into contact.
We note that the capture time is $\tau_0 = r_{\rm init}/((\alpha -1)v_t)$ if the target does 
not tumble at all ($d_0=0$). 
Results are shown in \textbf{Figure~\ref{fig:isotropic}}(a). The capture time $\tau$ approaches $\tau_0$ for 
small alert distances $d_0$, as the target allows the pursuer to come very close before tumbling, 
and is a decreasing function of $d_0$ for large $d_0$. 

It seems obvious 
that it is advantageous for the target to move away from the pursuer for evasion, which corresponds 
to the escape angle $\gamma = 0$, see Figure~\ref{fig:model}. 
Random isotropic tumbling, however, results in a complete misalignment, i.e., random redistribution 
of $\gamma$, and therefore a deviation from the desirable configuration for a target. 
On the other hand, erratic motion may additionally induce a misalignment of target and propulsion
direction also for the pursuer. Particularly after a tumbling event has occurred at a very short 
distance $r$, the pursuer may not be able to align with the new target direction quickly enough 
due to its limited maneuverability, which gives rise to large bearing angles 
$\langle \cos{\beta} \rangle \approx 0$,
analogously to the behavior of a pursuer chasing a fixed target, where a pursuer overshoots and 
circles around the target~\cite{goh:22,gass:23}. We therefore anticipate that an enhanced 
evasion may occur for $1 \lesssim d_0/\sigma \lesssim 1/\Omega_p$, where $1/\Omega_p$ implies a minimal 
distance for a significant probability for achieving a decrease in the bearing angle $\beta$. 
As shown in Figure~\ref{fig:isotropic}(a), the capture time indeed exhibits a transient behavior for 
intermediate $d_0$, where capture time becomes longer than 
$\tau_0$, in particular for $1/\Omega_p \gg 1$, which indicates that tumbling significantly 
enhances the escape rate. 

\subsection{Distance between target and pursuer}

In order to obtain a better understanding of the effect of tumbling on the escape dynamics, we examine the 
distribution of distances between pursuer and target in the {\em stationary limit}, i.e., we employ 
long-time trajectories without terminating the pursuit dynamics upon contact whenever averages 
of dynamic variables other than the capture time are considered.
As shown in Figure~\ref{fig:isotropic}(b), both for large and small $d_0$, the distance distributions exhibit 
a single peak at a short distance $r$. In the former case, frequent tumbling reduces the persistence length 
of target necessary for an escape, while in the latter case, the pursuer manages
to reach the target before tumbling is initiated. In sharp contrast, for intermediate alert distances 
$d_0$, a second peak appears at larger $r$, signalling the existence of a larger characteristic 
distance between pursuer and target. Visual inspection of trajectories indicates that this second peak corresponds 
to cases where the pursuer is forced to make a larger detour in order to reorient toward the target, see Movie S2, Supporting Information (SI). However, the 
values of $d_0$ where this characteristic distance is maximized are inconsistent with the optimal 
range of $d_0$ to maximize the capture time. 
The reason is that the detour giving rise to the increased average distance mostly occurs after the 
pursuer has come closer to the target than the contact distance $\sigma$, effectively missing the target.
The occurrence rate of such near misses might play an important role in the outcome of real-world pursuit with erratic maneuvers.
We revisit this issue in Section~\ref{sec:optimal} below.

\subsection{Run-length distribution}
\label{sec:iso_RLD}

A more appropriate quantity to characterize the dance of pursuer and target is the 
run-length distribution (RLD). To calculate this distribution, we initialize pursuer and target 
with the same orientation (along the connecting vector), with a distance $r_{\rm init}=20\sigma$. 
The results presented in Figure~\ref{fig:isotropic}(c) demonstrate that in the $d_0 \to \infty$ limit 
of distance-independent tumbling, 
RLDs display an exponential decay $e^{-\Omega_t (L/\sigma)} = e^{-C_t T}$ with persistence time $T$ 
or run-length $L$, which is the distribution of waiting times between events in a Poisson 
process with rate $C_t$.
As the alert distance $d_0$ decreases, power-law decays appear, indicating long 
time intervals of suppression of tumbling, where the target manages to maintain distances 
larger than $d_0$. As $d_0$ further decreases, RLDs then develop a peak at 
$L/\sigma \approx 20$ related to the turning radius $1/\Omega_p \approx 10$ of the pursuer,
and a fat tail for 
$L \gtrsim 100 \sigma$. To obtain an idea of the target behavior for the run-length 
at the peak and the fat tail, we extract the initial values of the escape and bearing angles, 
$\gamma_{\rm init}$ and $\beta_{\rm init}$, respectively, for each run sequence. 
Since the dynamics is deterministic apart from random tumbling, the initial configuration is decisive for 
the trajectory during a run sequence with $L \gg d_0$, where tumbling can only rarely happen. As depicted 
in Figure~\ref{fig:isotropic}(c), three motif (initial) configurations are identified, which are responsible 
for particular large run-lengths.  
While we find $\beta_{\rm init} \approx 0$ in all cases, 
$\gamma_{\rm init} \approx 0.4\pi$ is observed for run-lengths near the peak at $L/\sigma \approx 20$, where the target tumbles forward but somewhat off to the side.
Note that, while $\phi_t$ is uniformly distributed, 
$\gamma \equiv \phi_t - \theta +\pi$ is generally not.
In contrast, backward tumbling is responsible for the long run-lengths $L \gg d_0$, i.e., in the tails of
the distribution. In between, we observe 
configurations with $\gamma_{\rm init} \approx 0$, where the pursuer catches up with the target but
nearly misses (i.e. $r\gtrsim \sigma$) and then 
circles around to approach again. In this case, each run sequence can be separated into 
an initial approach with a short capture time and a second circling phase with a long capture time.

\subsection{Temporal autocorrelation function}
\label{sec:iso_autocorr}

We consider the temporal autocorrelation function of the target orientation from the initial 
conformation in the form
\begin{align}
C (\Delta t) = \langle {\bm e}_t (0) \cdot {\bm e}_t (\Delta t) \rangle.
\end{align}
As shown in Figure~\ref{fig:isotropic}(d) for $r_{\rm init}/\sigma=20$, the correlation function exhibits 
a negative minimum and negative long-time tail, which indicates that the presence of a pursuer leads to 
a rotational symmetry breaking, and
implies a preference for backward motion (toward the pursuer) at long times. First, the target 
keeps moving away from the pursuer, until the distance becomes sufficiently small, with 
$r \lesssim d_0$, which happens at $\Delta t \simeq (r_{\rm init}-d_0)/v_t$.
Then, the target tumbles frequently as long as the pursuer remains in the vicinity of the
target, in a short time window of duration $d_0/v_t$, until the distance from the pursuer increases to 
a large values $r> d_0$. The negative minimum signals that an occasional increase in distance occurs 
when the target tumbles backward by chance. After such a backward tumbling event, as the target escapes 
temporarily from the pursuer, tumbling is suppressed and the backward movement is maintained. 
Therefore, tumbling driven by the pursuer induces correlations in the tumbling direction, even though 
the tumbling is isotropic. This observation naturally implies a promising backward-tumbling strategy, 
where the target can achieve a large distance from the pursuer without a preceding close encounter.

\section{Anisotropic tumbling}
\label{sec:aniso}

\begin{figure*}
    \centering
    \includegraphics[width= \textwidth]{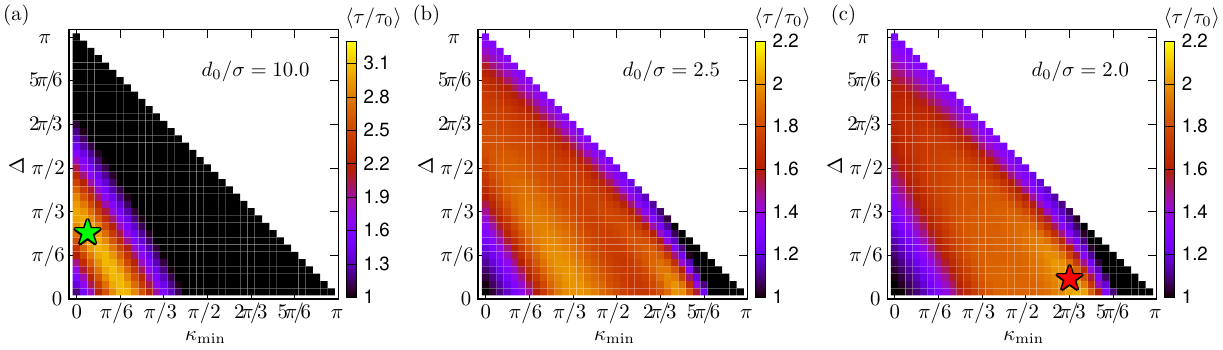}
    \caption{\label{fig:tumbling_angle}
    Capture time $\tau/\tau_0$ for various parameters of the tumbling-angle distribution. 
    (a) $d_0/\sigma=10.0$;
    (b) $d_0/\sigma=2.5$;
    (c) $d_0/\sigma=2.0$.
    Parameters are $\alpha=1.1$, $\Omega_p = 0.18$, $\Omega_t=0.5$, and $r_{\rm init}=20\sigma$. 
    The parameter values for forward and backward tumbling roughly correspond to 
    $\kappa_{\rm min} \lesssim \pi/6$, $\Delta \approx \pi/4$ (green star symbol in (a)) 
    and  $\kappa_{\rm min} \approx 2\pi/3$, $\Delta < \pi/4$ (red star symbol in (c)), 
    respectively.}
 \end{figure*}

The insight obtained from analysis of isotropic tumbling in Section~\ref{sec:iso_autocorr} 
suggests that a preferred tumbling direction can play an important role to avoid capture.
Accordingly, we investigate the capture time by considering  various   
tumbling-angle distributions. Results for the dependence of the capture time $\tau$ 
for three alert distances $d_0$ on the parameters $\kappa_{\rm min}$ and $\Delta$ of the 
tumbling-angle distribution are displayed in \textbf{Figure~\ref{fig:tumbling_angle}}, which demonstrate 
that the capture time depends critically 
on the tumbling-angle distribution. Overall, we identify two distinctive regimes, 
forward and backward tumbling, marked in Figure~\ref{fig:tumbling_angle}(a) and (c), 
respectively, which correspond to away and toward responses in terms of animal escapology~\cite{dome:11}. 
We note that the forward and backward regimes are not continuously connected, 
as demonstrated by two distinct capture-time peaks in Figure~\ref{fig:tumbling_angle}(b) 
for small $\Delta$.

\subsection{Forward tumbling}
\label{sec:forward}

\begin{figure*}
    \centering
    \includegraphics[width= 0.9\textwidth]{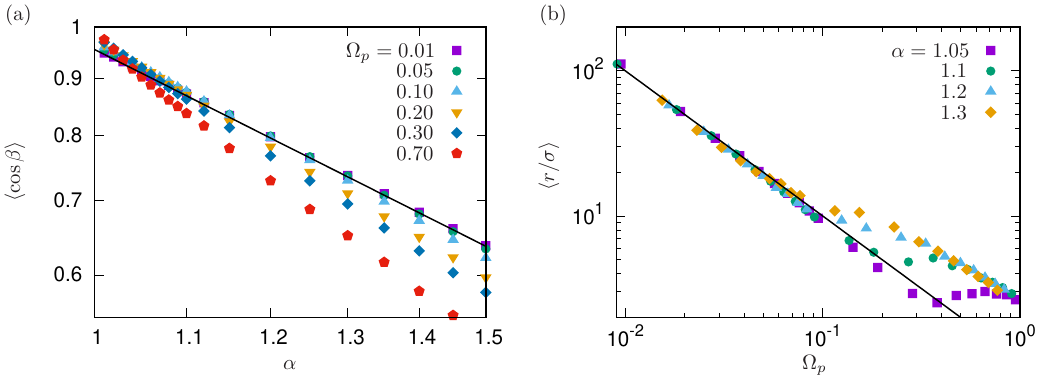}
    \caption{\label{fig:larged0}
    Forward tumbling. 
    (a) $\langle \cos{\beta} \rangle$ as a function of $\alpha$ for various $\Omega_p$, as indicated. 
    The solid line represents $\langle \cos{\gamma} \rangle/\alpha$ where 
    $\langle \cos{\gamma} \rangle = \pi/3$ for the used range of the tumbling angles, i.e., 
    $\kappa_{\rm min} = 0$, and $\Delta = \pi/6$.
    (b) Average pursuer-target distance $\langle r \rangle$ as a function of $\Omega_p$ for various
    speed ratios $\alpha$. The solid lines represents $1/\Omega_p$. 
    In both cases, $\kappa_{\rm min} = 0$, $\Delta = \pi/6$, $\Omega_t=0.5$, and $d_0=100\sigma$.
    }
\end{figure*}

For large alert distances $d_0$, where $\omega \approx 1$, and tumbling is independent of $r$, 
forward tumbling outperforms backward tumbling, enhancing the capture time significantly, as shown in 
Figure~\ref{fig:tumbling_angle}(a). Intuitively, this happens because a backward tumbling
at larges distances give the pursuer ample time to adapt its direction of motion, 
and therefore just contributes to reducing the distance $r$. 
For further theoretical analysis, we introduce a mapping of a RTP to an ABP 
(see Section~\ref{app:mapABP}), 
which allows us to rewrite the master equation, Equation~\ref{eq:tumbling_angle_W}, in a Langevin 
form in terms of the escape angle $\gamma$,
\begin{align} \label{eq:gamma}
\dot{\gamma} = -\frac{\sigma}{r}(\alpha \sin{\beta} + \sin{\gamma}) 
    +\sqrt{\frac{2\sigma}{l_p}}\eta_R,
\end{align}
which, together with Equation~\eqref{eq:beta} and~\eqref{eq:r}, describes the coupled dynamics of the pursuer and target. 
The target persistence length $l_p$ is given by Equation~\eqref{eq:lp}.

Equation~\eqref{eq:r} immediately implies that any deviation of $\gamma$ from zero decreases the 
capture time as long as $\beta$ remains small. Therefore, an escape by increasing the distance ($\dot{r}>0$) is only 
achievable if the bearing angle deviates significantly from zero due to the dynamics of $\gamma$ governed by Equation~\eqref{eq:gamma}.
This corresponds to the situation where a pursuer effectively fails to align with the target direction.
Such failures do not lead to overshoots by the pursuer in this case, 
as the persistent motion of the target is not significantly impaired during forward tumbling (see Movie S3, SI). 
Only for large $\alpha$, where the self-propulsion of the target becomes negligible, 
bearing angles are isotropically distributed, which yields $\langle \cos{\beta} \rangle \to 0$. 
Instead, the dynamics exhibits a simultaneous intermittent quasi-circular motion by both the target and the pursuer, where the pursuer moves on the outer circle with a larger radius, thereby traveling a longer distance than the target.
We note here that the path lengths of the target and pursuer are proportional to their speeds, 
such that the pursuer always travels a distance $\alpha$ times longer than that of its target.

A rough estimate of the quasi-stationary variables $r$ and $\beta$
for a quasi-circular part of the trajectory, with the assumption of nonzero $\beta$ (so that $|\sin{\beta}| \gg |\sin{\gamma}|$), yields 
$\langle \cos{\beta} \rangle \approx \langle \cos{\gamma} \rangle /\alpha$ and 
$\langle r/\sigma \rangle \approx 1 /\Omega_p$, 
in good agreement with the simulation results, as shown in \textbf{Figure~\ref{fig:larged0}}. 
Therefore, the trajectories of faster pursuers with smaller turning radii $\sigma/\Omega_p$ tend to deviate more easily 
from those of their targets than slower pursuers, 
as indicated by the fact that $\langle \cos{\beta} \rangle$ 
is a decreasing function of $\alpha$.
This approximation 
should be valid for small $\Omega_p$.
For $\Omega_p > \Omega_t$, we anticipate 
frequent captures and overshoots of the target, which lead to significant jumps in the escape angle 
$\gamma$ as well as the bearing angle $\beta$. As a result, the simulation results deviate from our analytical estimate at large $\Omega_p$, 
see Figure~\ref{fig:larged0}.



\subsection{Backward tumbling}

\begin{figure}
    \centering
    \includegraphics[width= 0.45\textwidth]{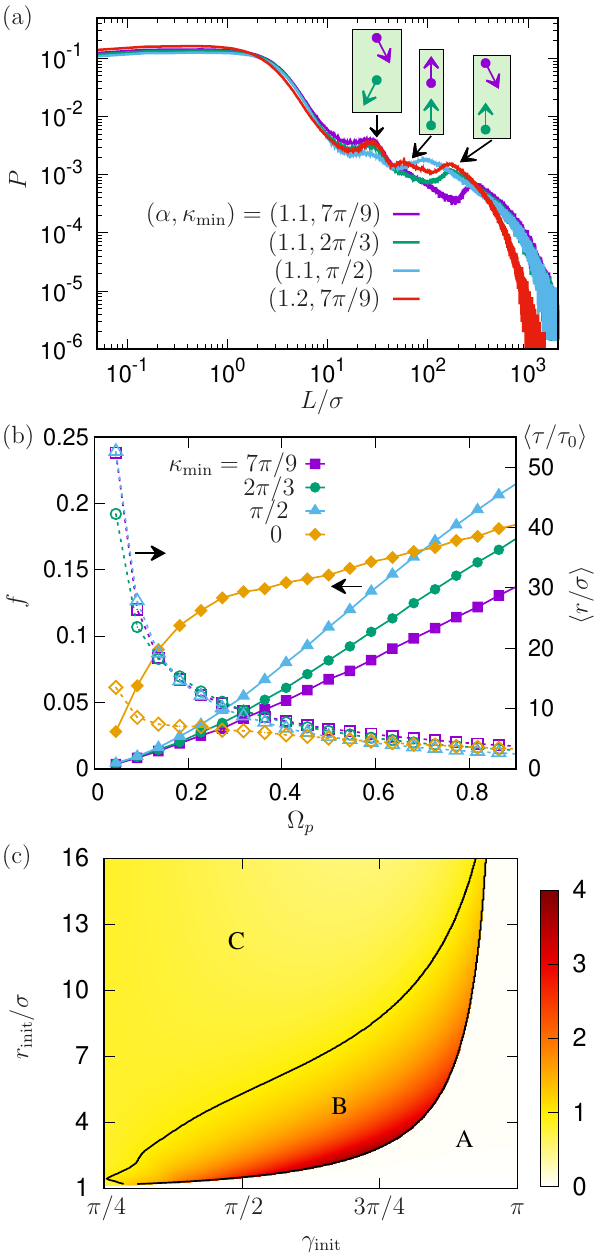}
    \caption{\label{fig:smalld0}
    Backward tumbling. 
    (a) Run-length distributions for $\Omega_t = 0.5$, $\Omega_p = 0.083$, $d_0/\sigma = 2.0$, and $\Delta=\pi/6$. Typical configurations 
    depicted as inset are extracted from the $(\alpha, \kappa_{\rm min}) = (1.2, 7\pi/9)$ data.
    (b) Tumbling frequency $f$ and average distance $\langle r/\sigma \rangle$ for small $d_0$. 
    Here, $\Omega_t=0.5$, $\alpha=1.1$, $d_0/\sigma = 2.0$, and $\Delta=\pi/6$. 
    (c) Capture time for the deterministic case (no tumbling), shown as a heat map. Region $B$ enclosed by a black solid line represents the cases where $\tau > \tau_0$. Here, $\alpha = 1.1$ and $\Omega_p=0.18$.}
\end{figure}

Backward tumbling outperforms forward tumbling for small alert distances $d_0$, see 
Figure~\ref{fig:tumbling_angle}(c). 
For backward tumbling, where $\kappa_{\rm min} >  \pi/2$, the mapping of RTPs to ABPs 
via Equation~\eqref{eq:approxP_forward} does not apply. 
Here, motif configurations provide insight into the dynamics, 
similar to the case of isotropic tumbling in Figure~\ref{fig:isotropic}(c).
As shown in \textbf{Figure~\ref{fig:smalld0}}(a), run-length distributions exhibit at most three peaks within 
the investigated ranges of parameters for backward tumbling. Apart from the aligned configuration with 
$\beta_{\rm init} \approx 0$ and $\gamma_{\rm init} \approx 0$, which has already 
been elucidated in Section~\ref{sec:iso_RLD}, we identify two other relevant motif configurations. 
For large $L$, we confirm that backward tumbling is responsible for the most probable escape 
configuration. In the case of intermediate run-lengths at $L/\sigma \approx 30$, we find 
$\beta_{\rm init} \approx -\pi$, indicating that the pursuer has passed by the target. Overall, 
only the former case of large $L$ is relevant for enhanced capture times, which corresponds to an 
escape scenario, where pursuer and evader persistently face in opposite directions, and simultaneously 
move away from each other.

Equation~\eqref{eq:r} indicates that the distance increase is maximized 
when $\beta = \pi$ and $\gamma = 0$, which then persist without disturbance by 
subsequent tumbling. Otherwise, another motion reversal by tumbling would result in a configuration 
favored by the pursuer. Tumbling-frequency data confirm that tumbling is indeed suppressed for large 
$\kappa_{\rm min}$, as shown in Figure~\ref{fig:smalld0}(b) (see also Movie S4, SI). Therefore, it is a combination of the 
restricted tumbling angle and small $d_0$ that enhances the escape rate for backward tumbling.

As further tumbling is suppressed in the case of (initial) backward tumbling, we consider a 
deterministic case, where the target moves in a persistent self-propulsion direction without (further) tumbling 
for simplicity. In this case, the resulting trajectory is completely predictable from 
the initial condition, which corresponds to the configuration right after a tumbling event with escape 
angle $\gamma_{\rm init}$. Without loss of generality, we consider the initial configuration where the 
target is located at the origin, and the pursuer
at ${\bm r}_p = (0,-r_{\rm init})^T$, corresponding to an initial separation $r_{\rm init}$.
We then numerically compute the capture time, varying the initial separation and escape angle. 
Here, we focus on the case where the pursuer are aligned along the connecting vector, i.e., $\beta_{\rm init}=0$. 
Accordingly, the capture time for a forward-tumbling target ($\gamma_{\rm init} = 0$) is 
$\tau_0 = r_{\rm init}/((\alpha-1)v_t)$. The resulting capture-time distribution for $\alpha=1.1$ 
and $\Omega_p = 0.2$ is presented in Figure~\ref{fig:smalld0}(b). The capture time $\tau$ is shortest for 
$\gamma_{\rm init} \approx \pi$, where the target is moving directly toward the pursuer (region A). 
With decreasing $\gamma_{\rm init}$, jumps in the capture time appear (region B), where the pursuer 
loses track of the target, while the target still manages to avoid a direct capture. The width of the 
region of $\tau > \tau_0$ is large for small $r_{\rm init}$, which emphasizes the conclusion that backward 
tumbling of the target is most effective at short distances. Moreover, region B mainly appears for 
$\gamma_{\rm init} \gtrsim \pi/2$, which corroborates the conclusion about the efficiency of the backward 
tumbling strategy for small $d_0$. As $\gamma_{\rm init}$ decreases further, the capture time gradually 
approaches $\tau_0$ (region C).

\subsection{Optimal tumbling direction}
\label{sec:optimal}

\begin{figure}
    \centering
    \includegraphics[width= 0.45\textwidth]{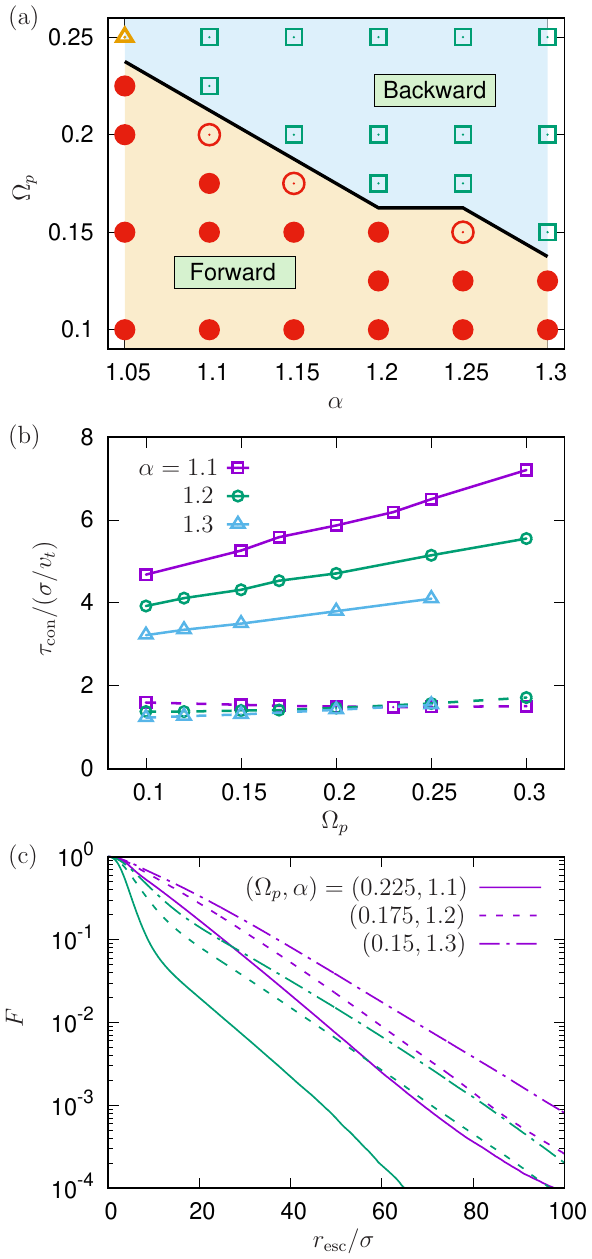}
    \caption{\label{fig:opt}
    (a) Optimal strategy maximizing the capture time between forward (large $d_0$) or backward (small $d_0$) 
    tumbling. Circles, squares, and triangles represent the parameter sets where forward, backward, and sideways 
    tumbling strategies, respectively, yield the minimum capture time. Filled circles indicate that the tumbling 
    increases the capture time to more than twice that of the deterministic case.
    (b) Contact duration time $\tau_{\rm con}$ as a function of pursuer maneuverability $\Omega_p$ for various 
    speed ratios $\alpha$ as indicated. Solid and dashed lines depict $\tau_{\rm con}$ for forward- and 
    backward-tumbling scenarios, respectively.
    (c) Survival function $F$ versus escape distance $r_{\rm esc}$, for various $\Omega_p$ and $\alpha$, 
    as indicated. Purple and green lines present backward and forward cases, respectively.
    In all cases, $\Omega_t = 0.4$ and $r_{\rm init}/\sigma=30$.
    }
\end{figure}

In order to draw conclusions for an ``optimal'' tumbling strategy,  we examine different tumbling-angle 
distributions. Specifically, we consider three 
tumbling-angle distributions which are designated as forward tumbling ($\kappa_{\rm min} = 0$), 
sideways tumbling ($\kappa_{\rm min}=\pi/4$), and backward tumbling ($\kappa_{\rm min} = \pi/2$), 
where in all cases, we choose $\Delta=\pi/4$. 

As a criterion of optimal escape strategy, we employ the
maximization of the capture time under variation of $d_0$ in all cases.
The resulting diagram in \textbf{Figure~\ref{fig:opt}}(a) displays the dependence of the best 
choice of tumbling angles on 
the pursuer maneuverability and speed. For a pursuer with low speed and small maneuverability, 
forward tumbling of the target results in maximal capture times, while for pursuer with high speed and large 
maneuverability, backward tumbling with small $d_0$, typically in the range $1.5 \leq d_0/\sigma \leq 10$, 
is preferable for the target. 
The same tendency is observed when $\Omega_t$ is varied, where for increasing or decreasing $\Omega_t$, 
forward or backward tumbling becomes advantageous, respectively. Therefore, we conclude that if the pursuer 
is less ``athletic'' a target may simply make frequent forward tumbling independent of the distance 
(large $d_0$). Surprisingly, the target still maintains rather persistently a relatively short distance 
from the pursuer, as obtained from the stationary solution of the dynamics, see Figure~\ref{fig:larged0}(b) 
and Section~\ref{sec:forward}. In this case, we observe a a significant increases in the capture time, e.g., 
exceeding twice the deterministic value $\tau_0$, and in some cases even hundreds of times $\tau_0$, except 
for the parameter values near the boundaries with the backward region, see Figure~\ref{fig:opt}(a). Therefore, 
exhausting pursuers by running away for a long time can be a plausible strategy for evaders in this case.

Such significant enhancements in capture time are observed only in the regime where 
$\alpha$ and $\Omega_p$ are small. In the regime of large $\alpha$ and $\Omega_p$, 
the increases in the capture time are limited, typically ranging from a few to several 
tens of percent for both forward- and backward-tumbling strategies. In other words, 
escaping from an athletic pursuer is very difficult. Backward-tumbling strategies 
outperform other strategies in these challenging scenarios.
The results for the bearing-angle distributions confirm this picture. Forward tumbling with small $\Omega_p$ and 
$\alpha$ exhibits a high peak at $\cos{\beta} =1$ (result not shown), implying that the pursuer predominantly 
follows the evader from behind, which suggests long, successful evading sequences. Otherwise, the bearing-angle 
distributions become broader, indicating frequent overshoots, and therefore frequent contacts, of the target 
by the pursuer.

Notably, a successful contact by the pursuer may not always lead to a capture. For instance, an abrupt change 
in speed due to a quick turning maneuver may induce an erratic response by the pursuer, causing the pursuer 
to miss its target, like in zebrafish predation~\cite{free:18}. In this context, backward-tumbling strategies 
provide additional advantages.

First, backward tumbling minimizes the contact duration in general. For quantification, we consider the average 
time interval $\tau_{\rm con}$ between the first capture ($r < \sigma$) and the time point when the target-pursuer 
distance again exceeds the capture distance ($r > \sigma$). As shown in Figure~\ref{fig:smalld0}(b), 
$\tau_{\rm con} > 3\sigma/v_t$ in forward-tumbling cases, whereas $\tau_{\rm con} < 2\sigma/v_t$ for backward 
tumbling; in the latter case, the motion of pursuer and target in opposite directions obviously facilitates 
rapid separation. Moreover, the values of $\tau_{\rm con}$ remain mostly unchanged as $\alpha$ and $\Omega_p$ 
increase, indicating that the contact duration will be short even for highly athletic pursuers.

Second, backward tumbling typically yields large target-pursuer distances. Instead of permanently outrunning 
the pursuer, a temporary escape with a significant distance could be enough to get out of sight of the pursuer, 
e.g., by hiding behind cover, in more complex, realistic environments.
In this case, pulling away from the pursuer, see e.g., fat tails in Figure~\ref{fig:isotropic}(b), 
can be a good indicator of escape. In this manner, we introduce the escape distance $r_{\rm esc}$, and 
calculate the complementary cumulative distribution, also called the survival function, 
$F(r_{\rm esc}) \equiv 1 - \int_{r_{\rm esc}}^\infty P(r') dr'$, which corresponds to the escape probability 
for given $r_{\rm esc}$. Figure~\ref{fig:opt}(c) presents survival functions $F$ for parameter values where 
backward tumbling is preferable in Figure~\ref{fig:opt}(a).
For both forward- and backward- tumbling strategies, the survival function exhibits exponential decay at 
large distances, indicating that overshoots following a contact lead to a temporary large separation between 
the pursuer and target, which can be quantified by the characteristic distance of the decay, in line with 
the exponential decay discussed in Ref.~\cite{goh:22} where a pursuer circles around a fixed target. In 
the forward-tumbling cases, however, additional quick drops of the survival function at small $r_{\rm esc}$ 
are conspicuous, suggesting that close approaches and subsequent overshoots rarely lead to significant 
increases in the survival probability. In stark contrast, such initial decays do not appear 
for backward-tumbling scenarios.


\section{Conclusion}

We have investigated the run-and-tumble escape strategy of a cognitive target from a 
self-steering pursuer with limited maneuverability. Our main conclusion is that a
hunted agent has to employ all information available for a successful escape strategy. This 
includes information about the instantaneous position of the predator. 
The decisions about the most promising escape moves, and their execution, have to be continuously 
adapted to the varying external conditions. In the system of run-and-tumble escape studied here,
this implies that the tumbling rate, and possibly the tumbling angles, 
have to be regulated and adapted to the pursuer behavior in order to achieve a preferable 
configurations for the target.

Frequent tumbling does not necessarily enhance the escape rate, 
because it leads to highly localized diffusive behavior, and therefore can actually be more advantageous 
for the pursuer than for the target.  When the pursuer is still far away, it is better for a target to tumble 
forward to induce a continuous disturbance in the bearing angle of pursuers
while not significantly affecting self-propulsion speed. Such a conclusion seems to support the 
``predator-avoidance'' hypothesis of the evolution of bipedal hopping of animals~\cite{mcgo:18}. 
However, if the pursuer is already close by, the target can take advantage of backward 
tumbling, which is momentarily risky but provides the chance to catch the pursuer on the wrong foot, 
and thus achieve temporarily a large 
head-start distance. Both strategies are used in real-world pursuit-evasion dynamics in biological
systems, e.g., see a penguin chased by whales~\cite{penguin_insta}, or prey chased by 
cheetahs \cite{wils:13}. A temporary large head-start distance might be sufficient for escape 
in a complex environment, where it may provide the possibility to disappear from the visual 
perception of the pursuer by going into hiding unnoticed.

The results can also be applied to steering strategies for autonomous robots and drones. For drones, 
a three-dimensional extension of our study would be particularly beneficial, while hydrodynamic 
effects need to be taken into account for microrobots in fluid environments~\cite{borr:22,goh:23}. 
Even though the development of robots capable of agile turning remains challenging, recent advances 
in the control of robot locomotion, especially combined with machine learning, are remarkable, 
particularly in quadruped robots~\cite{wang:11robot,yang:24}. In several cases, the developments 
of robotic controls and movement has already been inspired by animal locomotion~\cite{toua:12,pate:13,math:18}, 
demonstrating the importance of the fundamental link between robots and animals as intelligent 
active agents. In this context, our results provide guidance for the design and control of robotic 
systems. 

\appendix

\section{Mapping of forward tumbling RTP onto ABP}
\label{app:mapABP}

Assuming $\kappa_{\rm min}=0$, in polar coordinates comoving with the target, located at the origin, we rewrite the functional $\mathcal{W}$ in Equation~\eqref{eq:tumbling_angle_W} in the form
\begin{align}
\mathcal{W}[P] = \frac{1}{2\Delta} \int_{\theta-\Delta}^{\theta+\Delta} {\rm d}\phi_t P(\phi_t;t).
\end{align}
where $\gamma \ll 1$ has been assumed, which means 
that the target is moving away from the pursuer. If $P(\phi_t)$ and its derivatives are 
slowly-varying within the short interval $[\theta-\Delta,\theta+\Delta]$, we may further obtain
\begin{align} \label{eq:approxP_forward}
\frac{1}{2\Delta}\int_{\theta-\Delta}^{\theta+\Delta} {\rm d} \phi_t\,
P(\phi_t) \approx P(\phi_t)|_\theta + \frac{\Delta^2}{6} P''(\phi_t)|_\theta,
\end{align}
with which Equation~\eqref{eq:target_RTP} can be cast into a one-dimensional diffusion equation 
with the diffusion coefficient $D_{\rm FT} = \Omega_t \Delta^2 / 6$.
Here, the random tumbling is modeled by rotational diffusion of the target propulsion direction.
The diffusive motion can equivalently be described by the Langevin equation
\begin{align}
\dot{\phi_t} = \sqrt{2D_{\rm FT}} \eta_R, \label{eq:phit_eta}
\end{align}
where $\eta_R$ is Gaussian white noise with zero mean and $\langle \eta_R(t) \eta_R(t')\rangle = \delta(t-t')$.
Then the dynamics of the RTP can be mapped onto that of an ABP with persistence length 
\begin{align} \label{eq:lp}
\frac{l_p}{\sigma} \approx \frac{6}{\Omega_t\,\Delta^2}.
\end{align}

\end{document}